\documentstyle[aps,preprint]{revtex}
\def\d{\partial}
\def\<{\langle}
\def\>{\rangle}
\def\x{{\bf x}}
\firstfigfalse
\begin{document}
\preprint{CU-TP-1029}
\title{Light Goldstone boson and domain walls in the $K^0$-condensed
phase of high density quark matter}

\author{D. T. Son\footnote{email: son@phys.columbia.edu}}

\address{Physics Department, Columbia University, New York, New York
10027}

\address{and RIKEN-BNL Research Center, Brookhaven National
Laboratory, Upton, New York 11973}

\date{August 2001}

\maketitle

\begin{abstract}
It is pointed out that $K^0$ condensation in high density matter gives
rise to an extremely light Goldstone boson whose mass comes entirely
from weak interactions.  This implies the existence of metastable
non-topological domain walls with a long lifetime.  We comment on the
mass of the superfluid mode if baryon number is violated.

\end{abstract}
\tighten

Recently, many authors \cite{Schaefer,BedaqueSchaefer,KaplanReddy}
emphasize the likelihood of kaon condensation in the color-flavor
locked (CFL) \cite{CFL} phase of QCD at high baryon densities.  The
crucial observation is that kaons have small masses at high densities
\cite{inverse}.\footnote{Here and after, by ``kaons'' we mean the
quasiparticle excitations of the CFL phase which carry the same
quantum numbers as the kaons in vacuum.}  A relatively small
strangeness chemical potential is thus sufficient to drive kaon
condensation.  Moreover, it was argued that the mass of the strange
quark also works in favor of kaon condensation \cite{BedaqueSchaefer}.

In contrast to the conventional charge kaon condensation in nuclear
matter \cite{KaplanNelson}, in the CFL phase it is the neutral kaons
which are more likely to condense.  This is due to the inverse mass
ordering \cite{inverse} of mesons in the CFL phase, which makes the
neutral kaons lighter than the charge kaons, at least at very high
densities.  The fact that neutral kaons rather than charge kaons
condense might have important astrophysical consequences, since it
implies that the kaon-condensed phase does not require electrons to be
electrically neutral, which is one of the properties of the pure CFL
phase without kaon condensation \cite{RajagopalWilczek}.  In this
paper, we show that the $K^0$-condensed phase possesses another
distinct feature: it has in its spectrum an extremely light bosonic
particle, whose presence implies the existence of non-topological
metastable domain walls.

Let us first review the symmetry arguments underlying this feature.
Like most Bose-Einstein condensates, the $K^0$ condensate
spontaneously breaks a global U(1) symmetry.  The choice of the broken
generator is not unique: one can always add an unbroken generator to a
broken one.  The simplest operator is strangeness,
\begin{equation}
  S=\int\!d\x\, \bar s\gamma^0 s \, .
  \label{S}
\end{equation}
Since $K^0$ carries a strange charge, its condensation breaks the
corresponding symmetry.  From Goldstone's theorem, one expects a
Goldstone boson to appear in the spectrum.\footnote{If the isospin
symmetry was exact, there would be two Goldstone bosons, one with a
linear dispersion relation and another with a quadratic dispersion
relation \cite{MiranskyShovkovy,SSSTV}.  In this paper we will
consider the realistic case when the isospin symmetry is not exact.}
This boson is the U(1)$_S$ phase of the condensate, which will be
denoted as $\varphi$.  In addition, the system inherits the
spontaneous breaking of the baryon U(1) symmetry from the CFL quark
pairings.  It might appear that the $K^0$-condensed phase is a
two-component superfluid, the dynamics of which is determined by two
U(1) phases.

A closer examination reveals that the particle arising from the
U(1)$_S$ breaking is in fact only a {\em pseudo}-Goldstone boson.
Although strangeness is an exact symmetry of QCD, it is violated by
weak processes, hence the would-be Goldstone boson acquires a mass.
Since this mass comes entirely from weak interactions, it must be very
small, much smaller than other hadronic masses in the theory. It is
proportional to the square root of the Fermi constant $G_F$, in the
same way as the pion mass (in vacuum) is proportional to the square
roots of the quark masses which violate the conservation of the axial
currents.


The existence of a light Goldstone boson in the spectrum leads to the
appearance of metastable domain walls.  Indeed, at very low energies,
the system can be described in term of the variable $\varphi$ alone
[if one freezes the U(1) baryon phase].  The effective Lagrangian for
$\varphi$ must have the form
\begin{equation}
  L = {f^2\over2} [(\d_0\varphi)^2 - u^2 (\d_i\varphi)^2] - V(\varphi)
  \, ,
  \label{LV}
\end{equation}
where $f$ is the decay constant of the boson (which of order $\mu$),
$u$ is its velocity, and $V(\varphi)$ comes entirely from the explicit
violation of strangeness by weak interactions.  Due to the nature of
$\varphi$ as a phase variable, $V(\varphi)$ is required to be a
periodic function of $\varphi$.  Moreover, to leading order in $G_F$,
$V(\varphi)\sim\cos\varphi$.  The simplest way to see that is to
express the superfluid ground state with a definite value of $\varphi$
as a superposition of states with definite values of strangeness,
\begin{equation}
  |\varphi\> = \sum e^{iS\varphi} |S\> \, .
  \label{superfluid}
\end{equation}
To leading order in perturbation theory, the energy shift of the state
$|\varphi\>$ caused by an interaction Hamiltonian $H_{\rm int}$ is
$\<\varphi|H_{\rm int}|\varphi\>$.  The Hamiltonian of weak
interactions has only $\Delta S=1$ matrix elements, so $V(\varphi)$
is proportional to $\cos\varphi$, without higher harmonics.

The Lagrangian (\ref{LV}) now becomes that of the sine-Gordon model,
\begin{equation}
  L = {f^2\over2} [(\d_0\varphi)^2 - u^2 (\d_i\varphi)^2] 
      + f^2 m^2\cos\varphi \, ,
  \label{sG}
\end{equation}
where the coefficient in front of the $\cos\varphi$ term was written
in such a way that $m$ is the mass of the Goldstone boson.  It is well
known that the sine-Gordon theory possesses a domain wall solution,
which interpolates between $\varphi=0$ and $\varphi=2\pi$,
\begin{equation}
  \varphi(z) = 4 \arctan e^{mz/u} \, .
  \label{profile}
\end{equation}
Since the two values $\varphi(z=-\infty)=0$ and
$\varphi(z=+\infty)=2\pi$ actually refer to the same ground state,
this domain wall is non-topological and, in principle, can decay, but
its lifetime may be very long.  The domain wall here is an exact
replication, under different physical circumstances, of the axion
domain wall \cite{Vilenkin}, the U(1)$_{\rm A}$ domain wall of
high-density QCD \cite{SSZ}, and the domain wall of two-component
Bose-Einstein condensates of atomic gases \cite{BEC}.  Like in all
other cases, the appearance of the domain wall is deeply rooted in the
spontaneous breaking of an approximate U(1) symmetry.

The whole discussion above is based solely on symmetry arguments and
is independent of all details about the dynamics.  Therefore, we
should expect the light boson and the domain wall to be very robust
consequences of the $K^0$ condensation.  We now turn to the
calculation of the mass of the Goldstone boson in the kaon condensed
phase, from which we can extract the tension of the domain wall and
its lifetime.

The strangeness-violating piece of the four-fermion effective
Lagrangian for weak interactions is
\begin{equation}
  L = {4G_F\over\sqrt2}\cos\theta_c\sin\theta_c 
    (\bar s_L\gamma^\mu u_L)
    (\bar u_L \gamma_\mu d_L) + {\rm H.c.} 
\end{equation}
In order to generate a mass for the Goldstone boson we need an
effective interaction of the type $\bar s_R \bar u_R u_L d_L$ and
$\bar s_L \bar u_L u_R d_R$ (and their complex conjugates): these
terms, when averaged over the kaon-condensed state, lifts the
degeneracy of states with different $\varphi$ (see below).  Thus one
needs to transform two left-handed quarks into right-handed ones.
This can be done using two mass insertions.  One can put the mass
insertions in two different ways as in Fig.\ \ref{fig:vertex}.  When
all external lines are on mass shells, each of the mass insertions
introduces a factor of $\gamma^0m/2E$, where $E$ is the energy flowing
along the line where the mass insertion is made.  Since we will be
interested in the situation where all external momenta near the Fermi
surface, we can replace $E$ by $\mu$.  Thus we arrive to the following
effective interaction,
\begin{eqnarray}
  L_{\rm int} = {G_F \over \sqrt2\mu^2}\cos\theta_c\sin\theta_c 
  & & \, [\,  m_u m_s (\bar s_R\gamma^0\gamma^\mu u_L) 
  (\bar u_R\gamma^0\gamma_\mu d_L) \nonumber \\
  & &+\, m_u m_d (\bar s_L\gamma^0\gamma^\mu u_R)
    (\bar u_L\gamma^0\gamma_\mu d_R) \, ]
  \, + \, {\rm H.c.}
  \label{Lint}
\end{eqnarray}

\begin{center}
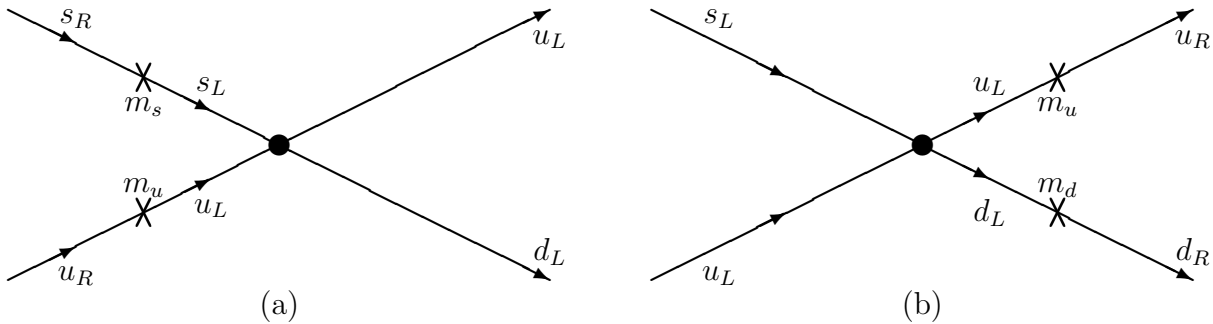
\begin{figure}
\setlength{\unitlength}{0.9mm}
\begin{picture}(175,50)(-40,-30)
\thicklines

\put(0,-25){\makebox(0,0)[b]{(a)}}
\put(95,-25){\makebox(0,0)[b]{(b)}}

\put(-40,-20){\vector(2,1){10}}
\put(-30,-15){\vector(2,1){20}}
\put(-10,-5){\vector(2,1){50}}

\put(-40,20){\vector(2,-1){10}}
\put(-30,15){\vector(2,-1){20}}
\put(-10,5){\vector(2,-1){50}}

\put(-21,-12){\line(1,2){2}}
\put(-21,-8){\line(1,-2){2}}
\put(-21,8){\line(1,2){2}}
\put(-21,12){\line(1,-2){2}}
\put(0,0){\circle*{3}}

\put(40,17){\makebox(0,0)[t]{$u_L$}}
\put(40,-17){\makebox(0,0)[b]{$d_L$}}
\put(-30,18){\makebox(0,0)[b]{$s_R$}}
\put(-10,8){\makebox(0,0)[b]{$s_L$}}
\put(-30,-18){\makebox(0,0)[t]{$u_R$}}
\put(-10,-8){\makebox(0,0)[t]{$u_L$}}
\put(-20,7){\makebox(0,0)[t]{$m_s$}}
\put(-20,-7){\makebox(0,0)[b]{$m_u$}}

\put(55,-20){\vector(2,1){20}}
\put(75,-10){\vector(2,1){30}}
\put(105,5){\vector(2,1){30}}

\put(55,20){\vector(2,-1){20}}
\put(75,10){\vector(2,-1){30}}
\put(105,-5){\vector(2,-1){30}}

\put(114,-12){\line(1,2){2}}
\put(114,-8){\line(1,-2){2}}
\put(114,8){\line(1,2){2}}
\put(114,12){\line(1,-2){2}}
\put(95,0){\circle*{3}}

\put(65,-18){\makebox(0,0)[t]{$u_L$}}
\put(65,18){\makebox(0,0)[b]{$s_L$}}
\put(135,17){\makebox(0,0)[t]{$u_R$}}
\put(135,-17){\makebox(0,0)[b]{$d_R$}}
\put(105,8){\makebox(0,0)[b]{$u_L$}}
\put(105,-8){\makebox(0,0)[t]{$d_L$}}
\put(115,7){\makebox(0,0)[t]{$m_u$}}
\put(115,-7){\makebox(0,0)[b]{$m_d$}}

\end{picture}
\caption{The effective vertices responsible for the mass of the
Goldstone boson}
\label{fig:vertex}
\end{figure}
\end{center}

For realistic quark masses, the second term in the square bracket of
Eq.\ (\ref{Lint}) can be neglected compared to the first term.  To
compute the average of $L_{\rm int}$ we need to be more specific about
the ground state.  The order parameters of the CFL phase of QCD are
two $3\times3$ matrices $X$ and $Y$, defined as
\cite{CasalbuoniGatto,inverse},
\begin{equation}
  \< q^{ai}_{L\alpha} q^{bj}_{L\beta}\>^* = 
  \epsilon_{\alpha\beta}\epsilon^{abc}\epsilon^{ijk}X^{ai} \,, \qquad
  \< q^{ai}_{R\dot\alpha} q^{bj}_{R\dot\beta}\>^* = 
  \epsilon_{\dot\alpha\dot\beta}\epsilon^{abc}\epsilon^{ijk}Y^{ai} \, ,
  \label{gap}
\end{equation}
where $a,b,c=1,2,3$ are color indices, $i,i,k=u,d,s$ is flavor
indices, and $\alpha,\beta,\dot\alpha,\dot\beta=1,2$ are Dirac spinor
indices.  The color-neutral order parameter describing the breaking of
chiral symmetry is the unitary matrix $\Sigma\sim X^\dagger Y$.  The
CFL ground state without kaon condensate corresponds to $\Sigma=1$,
while the $K^0$ condensate state corresponds to
\begin{equation}
  \Sigma = \left( 
  \begin{array}{ccc} 1 & 0 & 0 \\
   0 & \cos\theta & \sin\theta e^{i\varphi} \\
   0 & -\sin\theta e^{-i\varphi} & \cos\theta \end{array}\right) \, .
  \label{SigmaK}
\end{equation}
In Eq.\ (\ref{SigmaK}), $\theta$ is the parameter characterizing the
strength of kaon condensation (relative to chiral symmetry breaking),
which is fixed by the strangeness chemical potential and the kaon
mass, and $\varphi$ is our Goldstone phase variable which remains
undetermined if only strong interactions are taken into account.  We
will limit ourselves to the case of maximal $K^0$ condensation where
$\theta=\pi/2$.  This is achieved, for example, when the strangeness
chemical potential (or the effective chemical potential induced by the
strange quark mass \cite{BedaqueSchaefer}) is much larger than the kaon
mass.  The knowledge of $\Sigma$ does not fix $X$ and $Y$ due to the
freedom in performing a color rotation.  Since physical results should
not depend on the particular choice of $X$ and $Y$, we can choose, for
the simplicity of calculations, $X^{ai}=\delta^{ai}|X|$,
$Y^{ai}=\Sigma^{ai}|X|$.  With this choice, the $u_Ld_L$ diquark and
the $d_Rs_R$ diquark are of the third color,
\begin{equation}
  \< u_{L\alpha}^a d_{L\beta}^b \> 
    = \epsilon_{\alpha\beta}\epsilon^{ab3} |X|, \qquad
  \< u_{R\dot\alpha}^a s_{R\dot\beta}^b \> 
    = \epsilon_{\dot\alpha\dot\beta}\epsilon^{ab3} e^{i\varphi}|X| \, .
\end{equation}

Taking the average of $L_{\rm int}$ over the kaon-condensed state, we
find the potential term in the effective Lagrangian (\ref{LV}),
\begin{equation}
  V(\varphi) = -16 \sqrt2 G_F \cos\theta_c\sin\theta_c {m_u m_s\over \mu^2}
     |X|^2 \cos\varphi \, .
  \label{VX}
\end{equation}
In the CFL phase at asymptotically high densities \cite{SSZ}, 
\begin{equation}
  |X| = {3\over2\sqrt2} {\mu^2\Delta\over g} \, ,
  \label{X}
\end{equation}
where $\mu$ is the chemical potential, $\Delta$ is the
Bardeen-Cooper-Schrieffer gap (more precisely, the smaller one) of
color superconductivity.  Substituting Eq.\ (\ref{X}) to Eq.\
(\ref{VX}), we find
\begin{equation}
  V(\varphi) = -{18\sqrt2\over g^2}G_F\cos\theta_c\sin\theta_c
    m_u m_s \mu^2\Delta^2 \cos\varphi \, .
  \label{V}
\end{equation}
By comparing the kinetic term of Eq.\ (\ref{LV}) with that of the
non-linear Lagrangian for $\Sigma$, one finds that, for maximal $K^0$
condensation, $f$ is equal to the decay constant of the pseudoscalar
mesons, which has been computed in Ref.\ \cite{inverse},
\begin{equation}
  f^2 = {21-8\ln2\over18}{\mu^2\over2\pi^2} \, .
  \label{f}
\end{equation}
The velocity $u$ is also equal to that of pseudoscalar mesons in the
pure CFL phase, which is $1/\sqrt3$ \cite{inverse}.  From Eqs.\
(\ref{V}) and (\ref{f}), one find the mass of the Goldstone boson,
\begin{equation}
  m^2 = {162\sqrt2\pi\over21-8\ln2} {G_F\over\alpha_s}
   \cos\theta_c\sin\theta_c m_u m_s \Delta^2 \, .
  \label{m2}
\end{equation}
If one takes into account the second diagram (Fig.\ \ref{fig:vertex}b),
the factor $m_s$ in Eq.\ (\ref{m2}) will be replaced by $(m_s+m_d)$.

That the Goldstone boson becomes massless if $m_d=m_s=0$ can be seen
from a symmetry argument: the $K^0$ condensate is not neutral under
the following charge,
\begin{equation}
  Q^R_{s-d} = \int\!d\x\, (\bar s_R\gamma^0 s_R - 
     \bar d_R\gamma^0 d_R) \, ,
\end{equation}
which is an exact symmetry of both strong and weak interactions if
$m_d=m_s=0$.  Goldstone's theorem now guarantees the vanishing mass.

The origin of the factor $m_u$ in the mass formula is less clear:
there is no symmetry which implies that the Goldstone boson should
become massless when $m_u=0$.  Indeed, if one takes non-perturbative
effects into account, there is an instanton-induced effective vertex
$\bar u_R\bar d_R u_L d_L$ which can replace the two mass insertions
in Fig.\ \ref{fig:vertex}b.  This instanton contribution is
proportional to $m_s$ and does not vanish when $m_u$ goes to 0.
However, since the density of instantons quickly drops as the baryon
chemical potential is increased, their effects are likely to be small.

None of the uncertainties in our calculations should change the
qualitative conclusion that the mass of the boson is very small, which
is due to the presence of the Fermi constant $G_F$ in the mass formula
(\ref{m2}).  Quantitatively, if we substitute to Eq.\ (\ref{m2}) the
numerical values $\alpha_s=0.3$, $m_u=4$ MeV, $m_s=150$ MeV, and
$\Delta=100$ MeV, we obtain the estimate $m\sim 50$ keV.  It is
instructive to compare $m$ with the kaon masses found in Ref.\
\cite{inverse} for the pure CFL phase without kaon condensation.  The
ratio between the Goldstone boson mass to the kaon masses in the pure
CFL phase is
\begin{equation}
  {m^2\over m_K^2} = {\cal O}\biggl({G_F\mu^2\over\alpha_s}\biggr)
  \ll 1 \, .
\end{equation}

The large separation of mass scales between the Goldstone boson and
the other mesons in the theory is responsible for the existence and
metastability of the domain wall.  The situation is mathematically
similar to the cases previously considered in Ref.\
\cite{Vilenkin,SSZ} so we will present here only the final formulas,
and refer the reader to the literature for the details of the
calculations.  The domain wall tension can be computed from its
profile (\ref{profile}),
\begin{equation}
  \sigma = 8 u f^2 m \, .
\end{equation}
The spontaneous decay of the domain wall occurs via hole nucleation,
with a rate suppressed by the exponent of the corresponding bounce
solution \cite{Vilenkin,SSZ}
\begin{equation}
  \Gamma \sim \exp\biggl( -{16\pi\over3}{\nu^3\over u\sigma^2}\biggr)
  \, ,
\end{equation}
where $\nu$ is the tension of the vortex line which is the boundary of
the nucleated hole.  Its value is
\begin{equation}
  \nu = \pi u^2 f^2 \ln {m_K \over m} \, ,
\end{equation}
where by $m_K$ we denote the mass scale of other pseudoscalar mesons
in the theory.  Thus
\begin{equation}
  \Gamma \sim \exp\biggl( -{\pi^4u^3\over12} {f^2\over m^2}
   \ln^3{m_K\over m}\biggr) \, .
\end{equation}
Due to the large ratio $f^2/m^2\sim \mu^2/m^2$, the exponent in
$\Gamma$ is huge, so the domain wall is practically stable at zero
temperature with respect to hole nucleation.

In conclusion, we have shown that the $K^0$-condensed state of high
density matter possesses a unique type of Goldstone boson which owes
its existence to the dynamics of strong interactions, but its mass to
weak interactions.  The tiny mass of such a boson gives rise to
non-topological domain walls.  A network of such domain walls, in
principle, can be formed when the core of a neutron star (if it is to
form a $K^0$ condensate) cools to temperatures below the kaon
condensation temperature.  It would be interesting to explore the
implications of the presence of these domain walls in the cores of
neutron stars.

As a consequence of the nonzero mass of the Goldstone boson considered
in this paper, the only true gapless mode in the $K^0$-condensed phase
is the baryon U(1) phase.  Thus, as far as the macroscopic behavior is
concerned, both the CFL phases with and without $K^0$ condensate are
one-component superfluids at zero temperature.
Only at distance scales smaller than the $m^{-1}$ scale ($\sim10^{-9}$
cm which is large compared to the inter-quark spacing but still
microscopic) does the $K^0$-condensed system behave like a
two-component superfluid.

Finally, let us make a comment on the possibility of baryon number
violation.  In this case, the corresponding superfluid Goldstone mode
also acquires a mass.  Since the superfluid order parameter, in both
nuclear matter and quark matter, is a dibaryon ($nn$ in neutron
superfluid or $qqqqqq$ in the CFL phase), the mass square of the
superfluid Goldstone boson is proportional to the amplitudes of the
$\Delta B=2$ processes, but not the $\Delta B=1$ ones.  This can be
seen by making an expansion similar to Eq.\ (\ref{superfluid}).  The
$\Delta B=2$ interactions lead to neutron-antineutron oscillations
\cite{neutron-osc}.  We thus have the following crude estimate for the
mass of the superfluid mode,
\begin{equation}
  m^2 \sim m_P^3 \tau_{n\leftrightarrow\bar n} \, ,
\end{equation}
where $\tau_{n\leftrightarrow\bar n}$ is the characteristic time scale
for $n\bar n$ oscillations, and the proton mass $m_P$ was inserted for
dimensionality.  Using the experimental bound on $n\bar n$
oscillations, $\tau_{n\leftrightarrow\bar n}>10^8$ s \cite{PDG}, we
find $m<10^{-7}$ eV.  The thickness of the corresponding domain wall
is larger than about 1 m, and still might be less than the radius of
neutron stars.  However, unless the neutron star under consideration
rotates very slowly, a domain wall that thick is unlikely to exist
because of the high density of vortices (the mean distance between the
vortices is typically $10^{-2}$ cm \cite{Shapiro}).  In most grand
unified theories, the dominant baryon violating processes are $\Delta
B=1$. In these theories, the amplitudes of $\Delta B=2$ processes are
suppressed by $1/M_X^4$, where $M_X$ is at the GUT scale.  This makes
the mass of the superfluid mode very small, $m^2\lesssim m_P^6/m_X^4$.
For $M_X\sim10^{16}$ GeV, the Compton wavelength of the superfluid
mode is as large as 1 pc, so for all practical purposes it can be
considered massless.

I am indebted to E.J.~Weinberg for discussions.  I thank RIKEN,
Brookhaven National Laboratory, and U.S.\ Department of Energy
[DE-AC02-98CH10886] for providing the facilities essential for the
completion of this work.  This work is supported, in part, by a DOE
OJI grant.

\end{document}